\documentclass[prc,twocolumn,superscriptaddress,superscriptfootnote]{revtex4}
\usepackage{amssymb}

\usepackage{amsmath,amsfonts,amssymb,epsfig,color}

\begin{document}

\title{Staggering behavior of $0^+$ state energies in the $Sp(4)$
pairing model}
\author{K. D. Sviratcheva}
\affiliation{Department of Physics and Astronomy, Louisiana State University,
Baton Rouge, LA 70803, USA}
\author{A. I. Georgieva}
\affiliation{Department of Physics and Astronomy, Louisiana State University,
Baton Rouge, LA 70803, USA}
\affiliation{Institute of Nuclear Research and Nuclear Energy,
Bulgarian Academy
of Sciences, Sofia 1784, Bulgaria}
\author{J. P. Draayer}
\affiliation{Department of Physics and Astronomy, Louisiana State University,
Baton Rouge, LA 70803, USA}


\begin{abstract}
We explore, within the framework
of an algebraic $sp(4)$ shell model,  discrete approximations to various
derivatives of the energies of the lowest {\it isovector-paired}
$0^+$ states of atomic nuclei in the $40 \leq A \leq 100$ mass
range. The
results show that the symplectic  model can be used to successfully interpret
fine structure effects driven by the proton-neutron ($pn$)  and like-particle
isovector pairing interactions as well as interactions with higher
$J$ multipolarity. A  finite energy difference technique is used to investigate
two-proton and two-neutron separation  energies, observed irregularities found
around the $N=Z$ region, and the like-particle and $pn$ isovector
pairing gaps.
A prominent staggering behavior is observed between groups of even-even and
odd-odd  nuclides. An oscillation, in addition to that associated
with changes in
isospin values, that tracks with  alternating seniority quantum numbers related
to the isovector pairing interaction is also found.
\end{abstract}

\maketitle

\section{Introduction}

The observed staggering of energy levels in atomic nuclei requires a
theory that
goes beyond mean-field considerations \cite{BohrMottelson}. Staggering data
contain  detailed information about the properties of the nucleonic
interaction
and suggest the existence of high-order  correlations in the collective
dynamics. Most studies of staggering focus on two aspects of the  phenomena.
There are discrete angular momentum dependent oscillations of physical
observables; namely, of
$M1$ transitions in nuclei
\cite{Stefanova03} or of the energy levels themselves (e.g., in octupole
\cite{Phillips86,ChouCZ92,MDRRB99}, superdeformed
\cite{Flibotte93,Hamamoto94,Haslip98}, ground and $\gamma $
\cite{BohrMottelson,TokiW97,MDRRB00} bands in atomic nuclei, as well as in
molecular rotational bands \cite{BonatsosDDLMRR96}). And then there
are sawtooth
patterns of different physical quantities (most commonly binding energies) that
track with changes in the number of particles in a system (both in nuclei
\cite{Heisenberg32} and in  metallic clusters
\cite{deHeer93,Neukermans03}).

In nuclear structure physics, staggering behavior of the second type is
observed when one changes in a systematic way the usual nuclear
characteristics
such as proton ($Z$), neutron ($N$), mass ($A$) or isospin projection
($|Z-N|/2$)
numbers. Examples  of these nuclear phenomena include odd-even mass staggering
(OEMS)
\cite{BohrMottelson,MadlandN88MollerN92,SatulaDN98,AlhassidLN99,BenderRRM00,
DobaczewskiMNSS01,Duguet01,Velazquez02}, odd-even staggering in
isotope/isotone
shifts
\cite{Tomlinson62,Caurier01}, and zigzag patterns of the first
excited $2^+_1$
state energies in even-even nuclei \cite{DrenskaGM02}. The staggering
behavior of
a nuclear observable is most easily seen when discrete derivatives of
second or
higher order in its variable(s)  are considered. The aim of this approach is to
filter out the strong mean-field (global) effects  and in so doing
reveal weaker
specific features. In this way, for example, the OEMS, which is usually
attributed to the nuclear pairing correlations, manifests itself in  certain
finite differences of the binding energies that can provide for a
measure of the
empirical  pairing gap \cite{BohrMottelson}. Likewise, various discrete
approximation of derivatives (filters) of  the binding energies can
be considered
to investigate detailed properties of the  nuclear structure
\cite{GrossN83,GambhirRS83,ZhangCB89,ZamfirC91,KanekoH99,RopkeSSL00,Macchiavelli00,Vogel00}.

In this paper, we consider the binding energies of the $0^+$ ground
states of even-$A$ nuclei in the mass range $40\leq A \leq 100$, and for the
rest odd-odd nuclei with  a ($J^\pi\neq 0^+$) ground state, the energy of the
lowest  isobaric analog $0^+$ excited state  (which corresponds to the ground
state of the even-even neighbor). We refer to these states as lowest  {\it
isovector-paired} $0^+$ states \cite{SGD03}. Our aim is to investigate how
various,  comparatively small
but not insignificant, parts of the interaction between nucleons
influence these
states when  we consider higher-order discrete derivatives of their energies
within the framework of a  convenient systematics.

The algebraic pairing model \cite{SGGDI01,SGD03} we exploit is based on a
fermion realization of the symplectic $sp(4)$ algebra which is isomorphic to
$so(5)$ \cite{Ginocchio,Hecht,EngelLangankeVogelDobes}. It includes
an isovector
(isospin $T=1$) pairing interaction as well as a diagonal (in an isospin basis)
$pn$ isoscalar ($T=0$) part. The latter is proportional to a so-called
$T(T+1)$ symmetry
term\footnote{It is common to address the symmetry energy in a
slightly  different way: the $T(T+1)$ term together with the isospin dependence
of the isovector  pairing term yield symmetry ($\sim T^2$) and Wigner 
($\sim T$)
energies.}
\cite{SGD03}. The operators of the reduction $sp(4)\supset  u(2)\supset u(1)
\oplus su(2)$ provide for a convenient and useful classification of nuclei and
their corresponding ground and excited states. The systematics is in
terms of the
eigenvalues of these  operators, namely, the total number $n$ of
valence nucleons
and the third projection $i$ of the isospin and  their linear combinations.

We have already shown in \cite{SGD03} that the
$Sp(4)$\footnote{We distinguish between groups
denoted with capital letters [e.g. $Sp(4)$] and the associated algebras of
their generators denoted with lower-case letters [e.g. $sp(4)$].}
model leads to  a good reproduction of the experimental energies
\cite{AudiWapstra} of the lowest {\it isovector-paired}
$0^+$  state for even-$A$ nuclei, $32\leq A
\leq 100$. As pointed out \cite{SGD03}, although the $T=1$
like-particle pairing
energy and the $T=1$ $pn$ pairing energy yield $\triangle n=2$
staggering patterns
that are of  opposite phases, the total isovector pairing energy has a smooth
behavior. It is the symmetry term that  makes an accurate theoretical
prediction
of the regular zigzag pattern of the experimental energies in
isobaric sequences
possible. As a further and more detailed investigation, we now
consider different
types of  discrete derivatives of the Coulomb corrected \cite{RetamosaCaurier}
energy function according to the
$Sp(4)$ classification and with no adjustable parameters.

The symplectic $Sp(4)$ scheme not only allows for a systematic
investigation of
staggering patterns in the experimental energies of the even-$A$
nuclei, it also
offers a  simple algebraic model for interpreting the results. Moreover, this
detailed investigation  serves as a test for the validity and
reliability of the
$Sp(4)$ model and the interactions it includes.

\section{$Sp(4)$ classification scheme}

We start with a brief outline of the
algebraic approach \cite{SGGDI01} used to interpret phenomena that have been
observed experimentally and are  related to the isovector ($T=1$ pairing
correlations) and isoscalar interactions in nuclei. The
$sp(4)$ algebra is realized in terms of creation $c _{jm\sigma }^\dagger$ and
annihilation $c _{jm\sigma }$ fermion operators with the standard
anticommutation relations
$\{c _{jm\sigma },c _{j^{\prime  }m^{\prime }\sigma ^{\prime }}^{\dagger
}\}=\delta _{j,j^{\prime }}\delta _{m,m^{\prime }}\delta _{\sigma ,\sigma
^{\prime }},$ where these operators create (annihilate) a particle of type
$\sigma =\pm 1/2$ (proton/neutron) in a state of total angular momentum $j$
(half integer) with projection
$m$ in a finite space $2\Omega =\Sigma _j (2j+1)$. In addition to the number
operator
$\hat{N}=\hat{N}_{+1}+\hat{N}_{-1}$ and the third isospin projection
$T_0=(\hat{N}_{+1}-\hat{N}_{-1})/2$, the generators of $Sp(4)$ are
\begin{eqnarray} T_\pm &=& \frac{1}{\sqrt{2\Omega}} \sum_{jm} c^\dagger_{jm,\pm
1/2}  c_{jm,\mp 1/2},
\label{Tgen} \\
      A^{\dagger }_{\mu =\sigma+\sigma^{\prime}}&=&
\frac{1}{\sqrt{2\Omega (1+\delta_{\sigma\sigma ^{\prime}})}}
\sum_{jm} (-1)^{j-m}
c_{jm\sigma}^\dagger c_{j,-m,\sigma ^{\prime}}^\dagger,\nonumber \\
      A_\mu &=& (A^{\dagger }_\mu)^\dagger,
\label{ABgen}
\end{eqnarray} where $\hat{N}_{\pm 1}$ are the valence proton (neutron) number
operators,
$T_{0}$ and $T_{\pm}$ are the valence isospin operators, and the generators,
$A^{\dagger }_{0,+1,-1}$, create a proton-neutron ($pn$) pair, a proton-proton
($pp$) pair or a neutron-neutron $(nn)$ pair of total angular momentum
$J^{\pi}=0^+$ and isospin $T=1$. A totally symmetric finite space is spanned by
the basis vectors constructed as ($T=1$)-paired fermions
\begin{equation}
\left|n_{+1},n_{0},n_{-1}\right) =\left( A^{\dagger }_{+1}\right)
^{n_{+1}}\left(
A^{\dagger }_{0}\right) ^{n_{0}}\left( A^{\dagger }_{-1}\right) ^{n_{-1}}\left|
0\right\rangle ,
\label{GencsF}
\end{equation} where $n_{+1,0,-1}$ are the numbers of pairs of each kind, $pp$,
$pn$, $nn$, respectively, and $\left| 0\right\rangle$ denotes the vacuum state.
In the like-particle pairing limit, $n_0$ gives the number of protons
(neutrons) not coupled to $J=0$ $pp$ ($nn$) pairs and hence defines the
usual seniority quantum number \cite{Racah,Kerman}, $\nu _1=n_0$. On the other
hand, in the $pn$ pairing limit another seniority number,
$2\nu _0=2n_{+1}+2n_{-1}$, is recognized that counts the particles not coupled
in $J=0$ $pn$ pairs.
The dependence of $\nu _0$ on $\nu _1$ within a given nucleus allows one
to consider only $\nu _1$ in the analysis; specifically, for a system of $n$
valence particles with isospin projection $i=(Z-N)/2$, the fully paired states
(\ref{GencsF}) differ in their coupling mode as the seniority quantum number
$\nu _1$ ($\nu _0=n/2-\nu _1$) changes by $\pm 2$ ($\mp 2$).
\begin{table}[th]
\caption{Realizations of the $u^\mu(2)=u^\mu(1)\oplus su^\mu(2)$
subalgebras of $sp(4)$.}
\begin{tabular}{l|ccc}
\hline Symmetry ($\mu $)&
  \begin{tabular}{c}
  $u^{\mu}(1)$ \\
  ($C_{1}^{\mu }$)
  \end{tabular}
  &
   \begin{tabular}{c}
   Eigenvalues \\
   of $C_{1}^{\mu }$
   \end{tabular}
					& $su^\mu(2)$\\
\hline
Isospin ($T$)    & $\hat{N}$      & $n$     & $T_+, T_0, T_-$
\\
$pn$ pairs (0)   & $T_0$          & $i$     & $A^{\dagger } _0,\frac{1}{2}
\hat{N} - \Omega, A_0$
\\
$pp$ pairs ($+$) & $\hat{N}_{-1}$ & $N_{-1}$ & $A^{\dagger }_{+1},
\frac{1}{2} (\hat{N}_{+1}-\Omega),A_{+1}$
\\
$nn$ pairs ($-$) & $\hat{N}_{+1}$ & $N_{+1}$ & $A^{\dagger }_{-1},
\frac{1}{2} (\hat{N}_{-1}-\Omega),A_{-1}$
\\
\hline
\end{tabular}
\label{tab:su2}
\end{table}

As a dynamical symmetry, the $Sp(4)$ symplectic group describes
isovector pairing
correlations and isospin symmetry through the four different reduction chains
$Sp(4)\supset U^{\mu }(2)\supset U^{\mu }(1)\otimes SU^{\mu }(2)$ with
$\mu =T,0,\pm $ (Table  \ref{tab:su2}).
The first order invariant of  $u^{\mu }(2)$,
$C_{1}^{\mu =\{T,0,\pm  \}}=\{\hat{N},T_0,\hat{N}_{\mp 1}\}$, realizes the
$u^{\mu }(2)\supset su^{\mu }(2) $ reduction and reduces the finite
action space
into a direct sum of unitary irreducible representations (irreps) of $U^{\mu
}(2)$  ($\{ n,i,N_{\mp  1} \}$ multiplets). Within a multiplet the third
projection generator of $SU^\mu(2)$ (middle operator in the fourth column in
Table \ref{tab:su2}) further reduces the $U^\mu(2)$ representation to a vector
with fixed quantum numbers $(n,i)$, or alternatively $(N_{+1},N_{-1})$,
to which corresponds a given nucleus (a cell in Table
\ref{tab:classifCa}). In this way the dynamical
$Sp(4)$ symmetry provides for a natural classification scheme of  nuclei as
belonging to a single-$j$ level or a major shell (multi $j$), which are mapped
to  the algebraic multiplets. This classification also extends to the
corresponding ground and  excited states of the nuclei including their {\it
isovector-paired} $0^+$ states.
\begin{table}[th]
\caption {Classification scheme of even-$A$ nuclei in the $1f_{7/2}$
shell. The
shape of the table is symmetric with respect to the sign of $i$ and
$n-2\Omega $.
$\Delta n=2$  in each $i$ multiplet (columns),
$\Delta i=1$  in each $n$ multiplet (rows), $\Delta N_{\pm 1}=2$  in  each
$N_{\mp 1}$ multiplet (diagonals). The subsequent action of the $SU^\mu(2)$
generators  (shown in brackets) constructs the constituents in a given
$SU^\mu (2)$ multiplet ($\mu =T,0,\pm $).}
\begin{tabular}{|c|cc|c|cccc}
\hline
$n\backslash i$ & \multicolumn{1}{|c|}{2} & 1& 0 & -1 &
\multicolumn{1}{|c}{-2} &
\multicolumn{1}{|c}{-3} &
\multicolumn{1}{|c|}{-4} \\ \hline 0 & &  & $_{20}^{40}$Ca$_{20}$    &  &  &  &
\\
\cline{1-1}\cline{3-5} 2 & & \multicolumn{1}{|c|}{$_{22}^{42}$Ti$_{20}$} &
$_{21}^{42}$Sc$_{21}$ &
$_{20}^{42}$Ca$_{22}$ &
\multicolumn{1}{|c}{} &  &  \\
\cline{1-6} 4 & $\swarrow $ & \multicolumn{1}{|c|}{$_{23}^{44}$V$_{21}$} &
$_{22}^{44}$Ti$_{22}$ &
$_{21}^{44}$Sc$_{23}$ &
\multicolumn{1}{|c}{$_{20}^{44}$Ca$_{24}$} & \multicolumn{1}{|c}{} &  \\
\cline{1-7} 6 & $\cdots $ & \multicolumn{1}{|c|}{$_{24}^{46}$Cr$_{22}$} &
$_{23}^{46}$V$_{23}$ &
$_{22}^{46}$Ti$_{24}$ &
\multicolumn{1}{|c}{$_{21}^{46}$Sc$_{25}$} &
\multicolumn{1}{|c|}{$_{20}^{46}$Ca$_{26}$} &  \\
\hline 8 & $\leftarrow ^{ {\tiny (T_+)}}$  &
\multicolumn{1}{|c|}{$_{25}^{48}$Mn$_{23}$} &
$_{24}^{48}$Cr$_{24}$ & $_{23}^{48}$V$_{25}$ &
\multicolumn{1}{|c}{$_{22}^{48}$Ti$_{26}$} &
\multicolumn{1}{|c|}{$_{21}^{48}$Sc$_{27}$} &
\multicolumn{1}{|c|}{$_{20}^{48}$Ca$_{28}$}
\\
\hline 10 & \multicolumn{1}{|c|}{$\cdots $}
&\multicolumn{1}{|c|}{$_{26}^{50}$Fe$_{24}$} &
\multicolumn{1}{|c|}{$_{25}^{50}$Mn$_{25}$}
&\multicolumn{1}{|c|}{$_{24}^{50}$Cr$_{26}$}
&\multicolumn{1}{|c|}{$_{23}^{50}$V$_{27}$}
&\multicolumn{1}{|c|}{$_{22}^{50}$Ti$_{28}$} \\
\cline{1-7} 12 & \multicolumn{1}{|c|}{{\tiny $(A^{\dagger }_{-1})$} 
$\searrow $ }
&$\vdots $ &
$\downarrow$ {\tiny ($A^{\dagger }_0$)} &\multicolumn{1}{|c|}{$\vdots
$} & \multicolumn{1}{|c|}{$\swarrow $ {\tiny $(A^{\dagger }_{+1})$}} & \\
\cline{1-6}
\end{tabular}
\label{tab:classifCa}
\end{table}

The general model Hamiltonian with $Sp(4)$ dynamical symmetry, which
consists of one- and two-body  terms, can be expressed through the
$Sp(4)$ group generators,
\begin{eqnarray} H =&-GA^{\dagger }_{0}A_{0}-F(A^{\dagger }_{+1}
A_{+1}+A^{\dagger }_{-1}A_{-1})  -E\frac{( T ^2-\frac{3\hat{N}}{4 })}{2\Omega}
\nonumber \\
&-C\frac{\hat{N}(\hat{N}-1)}{2}-(D-\frac{E}{2\Omega })
(T _{0}^2-\frac{\hat{N}}{4})-\epsilon  \hat{N},
\label{clH}
\end{eqnarray}
where $G,F,E,C$ and $D$ are strength parameters  and  $\epsilon >0$
is the Fermi level energy.
This Hamiltonian conserves the number of particles and
the third isospin projection and changes the seniority quantum number
$\nu _1$ by zero or $\pm 2$, the latter implies scattering of a $pp$ pair and
a $nn$ pair into two $pn$ pairs and vice versa. The isospin breaking
Hamiltonian (\ref{clH}) includes an isovector ($T=1$) pairing interaction
($G\geq  0,F\geq 0$ for attraction) and a diagonal  (in an isospin basis)
isoscalar ($T=0$) force, which is related to a symmetry term ($E$). Within
a shell (a single-$j$ level, $1f_{7/2}$, or a major shell,
$1f_{(\frac{5}{2})}2p_{(\frac{1}{2},\frac{3}{2})}1g_{(\frac{9}{2})}$), a
reasonable estimate for the parameters in the Hamiltonian (\ref{clH}) 
is obtained
in a fitting procedure of the maximum eigenvalues of $|H|$, $E_0$, to the
Coulomb corrected \cite{RetamosaCaurier} experimental energies
\cite{AudiWapstra} of the lowest {\it isovector-paired} $0^+$ states in
even-$A$ nuclei \cite{SGD03}. Although the fits yield quite good agreement
with the relevant experimental values, a reproduction of the fine properties of
nuclear structure (typically of an order of magnitude or two less than the
energies that were fit) is not guaranteed due to the strong mean-field
contribution. For the  present investigation
these parameters in the energy operator (\ref{clH}) are not varied; their
values are fixed as:
$G=0.53\Omega$, $F=0.45\Omega$, $C=0.47$, $D=-0.97$, $E=-1.12(2\Omega)$,
$\epsilon=9.36$ in MeV for the $1f_{7/2}$ level  (with a $^{40}Ca$ core) and
$G=0.35\Omega$, $F=0.30\Omega$, $C=0.19$, $D=-0.80$,
$E=-0.49(2\Omega)$, $\epsilon=9.57$ in MeV for the
$1f_{(\frac{5}{2})}2p_{(\frac{1}{2},\frac{3}{2})}1g_{(\frac{9}{2})}$  shell
(with a $^{56}Ni$ core) \cite{SGD03}.
In the second case, the parameters of the effective interaction in the $Sp(4)$
model with degenerate multi-$j$ levels are likely to be influenced by the
nondegeneracy of the orbits. Nevertheless, as the dynamical symmetry 
properties
of the two-body interaction in nuclei from this region are not lost, the model
remains a good multi-$j$ approximation \cite{SGD03} and the extent to which it
provides for a realistic description can be further tested with the use of
various discrete derivatives of the energy function.

\section{Discrete derivatives and fine structure effects}
The symplectic $Sp(4)$ model [namely, the $E_0$ maximal eigenvalues of
$|H|$ (\ref{clH})] reproduces the Coulomb corrected energies of the {\it
isovector-paired} $0^{+}$ states quite well \cite{SGD03}. A more detailed
investigation  and a significant test for the theory is achieved through the
discrete approximation of the $\partial ^m E_0/\partial x^m$ derivatives
of the $E_0$ energy function,
\begin{eqnarray}
Stg_{\delta }^{(m)}(x)
=&\frac{Stg_{\delta}^{(m-1)}(x+\frac{\delta  }{2})-Stg_{\delta
}^{(m-1)}(x-\frac{\delta }{2})}{\delta },\ m\geq 2, \nonumber \\
Stg_{\delta }^{(1)}(x)=&\left\{
\begin{array}{c}
\frac{E_0(x+\frac{\delta }{2})-E_0(x-\frac{\delta }{2})}{\delta },\ 
m\text{-even}
\\
\frac{E_0(x+\delta )-E_0(x)}{\delta },\ m\text{-odd,}
\end{array}
\right.
\label{Stag_m}
\end{eqnarray}
expressed recursively with two terminating conditions depending on
the order $m$ (even or odd) of the derivative, where the variable is
$x=\{n,i,N_{+1},N_{-1}\}$  according to the
$Sp(4)$ classification and $\delta \ge 1$ is a discrete integer step. 
The present investigation is focused predominantly on the $\delta =1 \text{ or }
2$ cases [the way the different variables, $n,i,N_{+1}$ or $N_{-1}$, vary in the $Sp(4)$
systematics can be recognized in Table \ref{tab:classifCa}]. 

The first $(m=1)$ discrete derivative
defined in (\ref{Stag_m}), $(E_0(x+\delta )-E_0(x))/\delta $, is
related to the $\delta $-particle separation  energy, when $x$ counts
the (total, proton or neutron) number of particles. The general
$Stg_{\delta }^{(m)}(x)$ quantity represents a finite difference between the
$E_0$ energies of neighboring nuclei, for example
\begin{eqnarray}
&Stg_{\delta }^{(2)}(x)=\frac{E_0(x+\delta )-2E_0(x)+E_0(x-\delta 
)}{\delta ^2},
\end{eqnarray}
when $m=2$, and
\begin{eqnarray}
&Stg_{\delta }^{(3)}(x)=\frac{E_0(x+2\delta )-3E_0(x+\delta
)+3E_0(x)-E_0(x-\delta )}{\delta ^3},
\end{eqnarray}
when $m=3$, and filters out contributions to $E_0$ proportional to
$x^{m-1}$.

The filters (\ref{Stag_m}) are $(m+1)$-point expressions that account for
deviations from the common behavior of neighboring 
nuclei. When $m\ge 3$ the
$Stg_{\delta }^{(m)}(x)$ discrete derivative is independent of
strong mean-field effects, strictly speaking it cancels out all
regularly-varying linear and quadratic in $x$ contributions to the energy, that
are typically large, and only can provide for a description of
higher-order terms in the variable $x$, as well as for  discontinuities in the
energy function. In this way, the 
finite energy
difference isolates specific parts of the interaction that are comparatively
smaller and may vary substantially from one nucleus to its neighbors. 
While these
interactions do not contribute much to the overall trend of the $E_0$
energies, they play a very significant role in determining nuclear structure
properties.

The mixed derivatives also provide useful information about the  nuclear fine
structure effects and are defined as
\begin{eqnarray} &Stg _{\delta _1,\delta _2}^{(2)}(x,y)= \nonumber \\
=&\frac{E_0(x+\delta _1,y+\delta _2)-E_0(x+\delta _1,y)-E_0(x,y+\delta
_2)+E_0(x,y)}{\delta _1 \delta _2}
\label{Stag_2_mixed}
\end{eqnarray} where the variables represent quantities among the set
$(x,y)=\{n,i,N_{+1},N_{-1}\}$ and $\delta _{1,2} \ge 1$ is a discrete
increment in accordance with the $Sp(4)$ classification scheme (Table
\ref{tab:classifCa}).

Different types of discrete derivatives are considered and various  staggering
patterns are investigated in the following sections. The
corresponding components
of the interaction isolated through the energy difference filters can be
explained in analogous ways as in \cite{ZhangCB89,ZamfirC91}, in
addition to the
advantage  that because they are free of Coulomb effect they reflect phenomena
related only to nuclear forces.

\subsection{Discrete derivatives with respect to $N_{+1} $ and $ N_{-1}$:
the $N=Z$ region}

\begin{figure}[th]
\centerline{\epsfxsize=3.5in\epsfbox{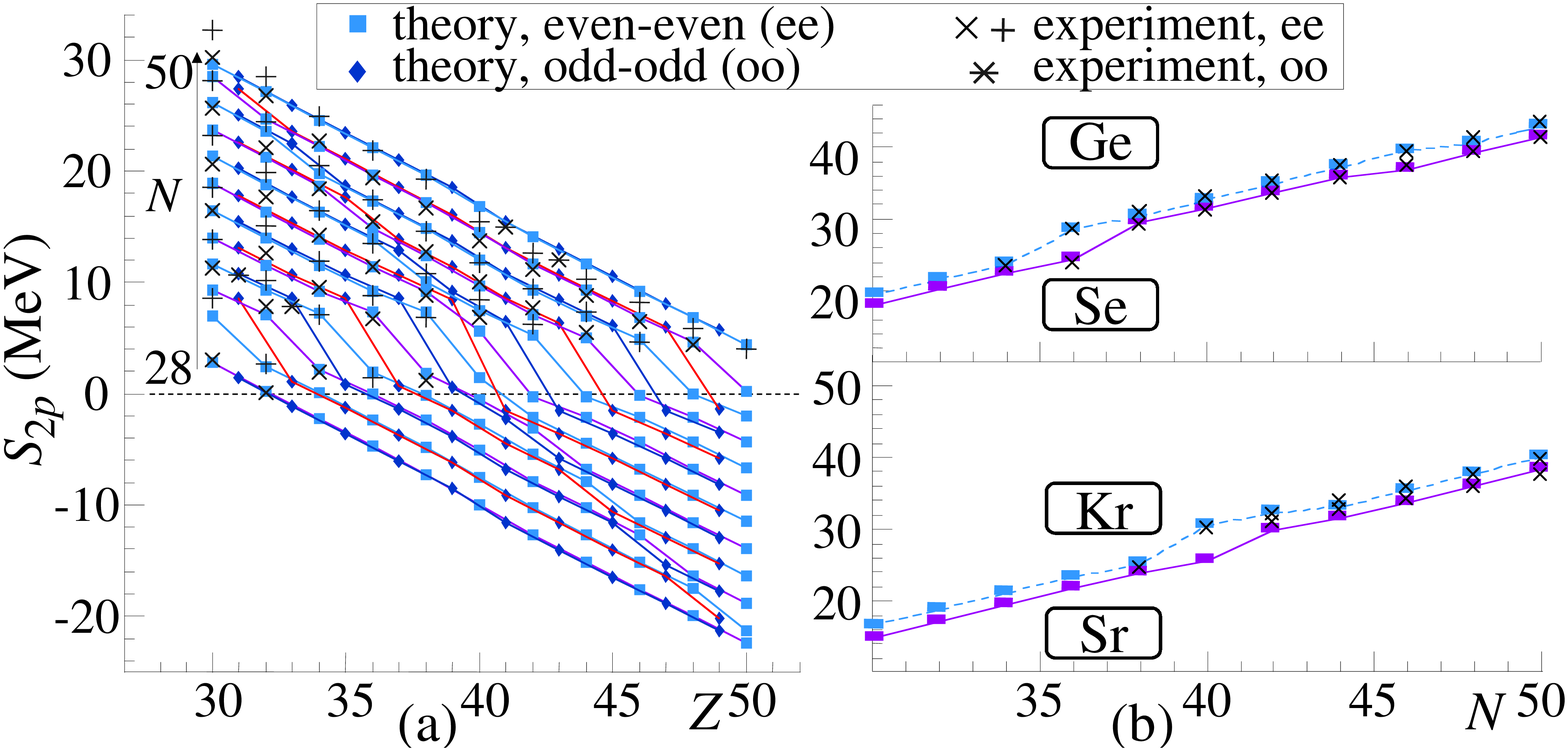}}
\caption{(Color online) The $S_{2p}$ two-proton separation energy in MeV for the even-even
(ee) and odd-odd (oo) nuclei in the
$1f_{(\frac{5}{2})}2p_{(\frac{1}{2},\frac{3}{2})}1g_{(\frac{9}{2})}$  major
shell: (a) versus number of protons for different isotones ($N=28-50$) (the
Coulomb repulsion  energy is taken into account); (b) versus number of neutrons
for Ge, Se, Kr, Sr isotopes.}
\label{NiS2p}
\end{figure}

For even-even nuclei, the discrete approximation of the $\partial 
E_0^C/\partial
N_{\pm 1}$ first derivative of the binding energies (including the Coulomb
repulsion  energy) is related to the well-known two-proton (two-neutron)
separation energy, which is  usually defined as
$S_{2p(2n)}(N_{\pm 1})=E_0^C(N_{\pm 1})-E_0^C(N_{\pm 1}-2)$ [see
Figure~\ref{NiS2p}(a) for a relation to proton number and Figure~\ref{NiS2p}(b)
for the difference of the  Coulomb corrected energies $E_0$ versus neutron
number]. The $Sp(4)$ theory reproduces very well the  available experimental
data \cite{AudiWapstra} (shown as `$\times $' or `$+$' symbols for
even-even nuclei and as $`\times \text{'} \hspace{-0.16in} -\ $ for odd-odd
nuclei in Figure~\ref{NiS2p}), especially the irregularity at $N_{+1}=N_{-1}$.
The zero point of $S_{2p}$ along an isotone sequence determines the
two-proton-drip line (dashed black line in Figure~\ref{NiS2p}), which
according to the $Sp(4)$ model for the
$1f_{(\frac{5}{2})}2p_{(\frac{1}{2},\frac{3}{2})}1g_{(\frac{9}{2})}$ major
shell lies near the following even-even nuclei:
\begin{equation}
\begin{tabular}{ccccc}
$^{60}$Ge$_{28}$, & $^{64}$Se$_{30}$, & $^{68}$Kr$_{32}$, & $^{72}$Sr$_{34}$,
& $^{76}$Zr$_{36}$, \label{dripLine}\\
$^{78}$Zr$_{38}$, & $^{82}$Mo$_{40}$, & $^{86}$Ru$_{42}$, & $^{90}$Pd$_{44}$,
& $^{94}$Cd$_{46}$,
\end{tabular}
\end{equation}
beyond which the higher-$Z$ isotones are unstable with respect 
to diproton
emissions. These nuclei are not yet explored as seen in Figure~\ref{NiS2p}
and an experimental comparison for the two-proton-drip line is expected to be
soon possible due to radioactive beam experiments near the limits of stability.
Yet, the findings of our model are in close agreement with the results of
other theoretical predictions
\cite{SmolanczukD93,Ormand97,MollerNK97,BrownCSV02}. Particularly,
the estimate for the two-proton separation energies in
\cite{Ormand97,MollerNK97,BrownCSV02} confirms the division in
nuclides such that the isotones with lower/higher $Z$ values than the nuclei
in (\ref{dripLine}) have positive/negative
$S_{2p}$ energies (compare to Figure~\ref{NiS2p}). In addition, the
two-proton separation energies for those of the nuclei in (\ref{dripLine})
considered also in the other studies are close in their estimates: 
the quadratic
mean of the difference in $S_{2p}$ between our model and \cite{Ormand97}
is $0.32$ MeV (in a comparison of the first three nuclei in (\ref{dripLine})),
is $0.78$ MeV when all the nuclei in (\ref{dripLine}) are
compared to \cite{MollerNK97} and is 0.43 MeV in a comparison to
\cite{BrownCSV02} of the first four nuclei in (\ref{dripLine}).
For odd-odd nuclei the zero point of
$S_{2p}$ can be also determined ($^{60}Ga_{29}$, $^{64}As_{31}$,
$^{68}Br_{33}$, $^{72}Rb_{35}$,
$^{76}Y_{37}$, $^{78}Y_{39}$, $^{82}Nb_{41}$,
$^{86}Tc_{43}$, $^{90}Rh_{45}$, $^{94}Ag_{47}$) although it does not
define the
drip line, as
$S_{2p}$ is a relation of the lowest $0^+$ state energies $E_0$  rather than of
the binding energies for most odd-odd nuclei.

\begin{figure}[th]
\centerline{\hbox{\epsfig{figure=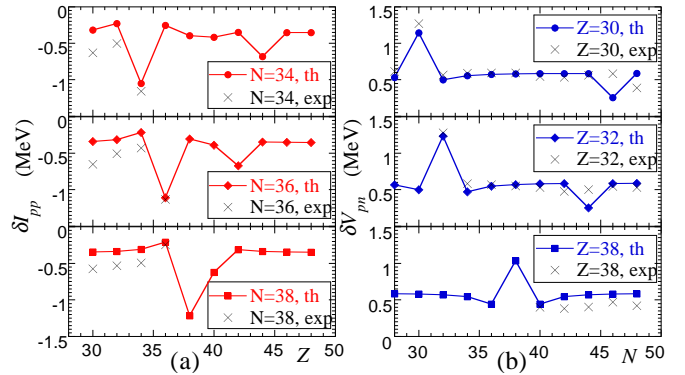,width=8.7cm}}}
\caption{(Color online) Second discrete derivatives of the $E_0$ energy
($1f_{(\frac{5}{2})}2p_{(\frac{1}{2},\frac{3}{2})}1g_{(\frac{9}{2})}$
shell): (a)
with respect to
$N_{+1}$, $\delta I_{pp}(N_{+1})$, as an estimation of the
nonpairing
like-particle nuclear interaction in MeV for the
$N=34,36,38$ multiplets; (b)
with respect to $N_{+1}$ and $N_{-1}$,
$\delta V_{pn}(N_{+1},N_{-1})$, for Zn, Ge, Sr isotopes.}
\label{NiDer2NpVpn}
\end{figure}

As a whole, the higher-order derivatives with respect to proton
(neutron) number
have a smooth behavior. This is because these derivatives reflect changes  only
within a sequence of either even-even or odd-odd nuclei. The discretization of
the
$\partial ^2 E_0/\partial N_{\pm 1}^2$ second-order derivative, $4\delta
I_{pp(nn)}(N_{\pm 1})=E_0(N_{\pm 1}+2)-2E_0(N_{\pm 1})+E_0(N_{\pm 1}-2)$
[$=4Stg_2^{(2)}(N_{\pm 1})$ (\ref{Stag_m}) ],  accounts for the interaction
between the last two $pp$ ($nn$) pairs in the ($N_{\pm 1}+2$) nucleus (Figure
\ref{NiDer2NpVpn}(a)). The average interaction $\delta I_{pp(nn)}$
may be used as an alternative way to \cite{ZamfirC91} to estimate the
nonpairing like-particle interaction
\footnote{The meaning of ``nonpairing'' relates to
$J\neq 0$ and $T\neq 1$ interaction or any interaction that is
different than the isovector pairing.  Also, here the approximation is of
$O(1/\Omega  )$.}  [of the
last two protons (neutrons)]. It shows no outlined staggering  pattern but a
repulsive peak around the
$N=Z$ nuclei in very good agreement with the experiment \cite{AudiWapstra} and with the 
results and discussions of
\cite{ZamfirC91}. Another smaller peak is observed around midshell  (Figure
\ref{NiDer2NpVpn}(a)), which is due to the particle-hole
discontinuity introduced in the  pairing theory. The analysis yields that as a
whole the $Sp(4)$ model reproduces the fine structure effects  in
interactions isolated via the $Stg_2^{(2)}(N_{\pm 1})$ filters.

\begin{figure}[th]
\centerline{\hbox{\epsfig{figure=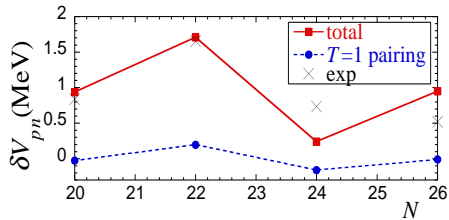,width=6cm,height=3cm}}}
\caption{(Color online) $\delta V_{pn}$ in MeV of the total binding energy ({\tiny
$\blacksquare $}) and of the $T=1$ pairing energy ($\bullet $) in comparison to experiment
($\times $) for Ti-isotopes in the $1f_{7/2}$ shell. The isovector  pairing interaction is
not enough to reproduce the experimental peak at $N=Z$.}
\label{Ca_cmp}
\end{figure}

Another aspect of the nuclear interaction is revealed by the  second-order
discrete mixed derivative of the energy \cite{JaneckeB74},
$\delta
V_{pn}(N_{+1},N_{-1})=(E_0(N_{+1}+2,N_{-1}+2)-E_0(N_{+1}+2,N_{-1})-E_0(N_{+1},N_{-1}+2)+E_0(N_{+1},N_{-1}))/4$
(\ref{Stag_2_mixed}). For even-even nuclei it was found to represent
the residual
interaction between the last proton and the last neutron
\cite{ZhangCB89,Brenner90} and it was empirically approximated by $40/A$
\cite{KanekoH99}.  The theoretical discrete derivative
(Figure~\ref{NiDer2NpVpn}(b)) agrees remarkably well with the  experiment
\cite{AudiWapstra}, especially in reproducing the typical behavior at $N_{+1}=N_{-1}$,
and is consistent
with the empirical  trend (on average , $\sim 0.71$ for $1f_{7/2}$ and $\sim
0.52$ for the major shell above the
$^{56}Ni$ core). It is well-known that the attractive peak in the
self-conjugate
nuclei cannot be described by a model with an isovector interaction only
\cite{Brenner90} and in this respect our model achieves this result  due to the
additional terms included in the Hamiltonian, mainly the symmetry term (Figure
\ref{Ca_cmp}). The $\delta V_{pn}$ energy difference provides for a  powerful
test for the symplectic model: the theory not only gives a thorough description
of  the isovector $pn$ and like-particle pairing but additionally accounts for
$J>0$  components of the $pn$ interaction in a consistent way with the
experiment. As a result the model can be used to provide for a reasonable
prediction of $\delta V_{pn}$ of proton-rich exotic nuclei as well as odd-odd
nuclei.

\subsection{Discrete derivatives with respect to $n$ and $i$: staggering
behavior
\label{DerStg}}

The $Sp(4)$ classification scheme can also be used to investigate  energy
differences with respect to the total number of particles $n$ and their
isospin  projection
$i$. Indeed, in contrast with the typical smooth behavior observed
for discrete
derivatives  with respect to $N_{+1}$ and
$N_{-1}$ that was highlighted in the preceding section, the
derivatives  with respect
to
$n$ and $i$ are the ones that reveal distinct staggering effects. They give a
relation  between even-even ($ee$) and odd-odd ($oo$)  nuclei and the  patterns
can be referred as an ``$ee-oo$" staggering.

\subsubsection{Second and higher-order derivatives in one variable}
\begin{figure}[th]
\centerline{\hbox{\epsfig{figure=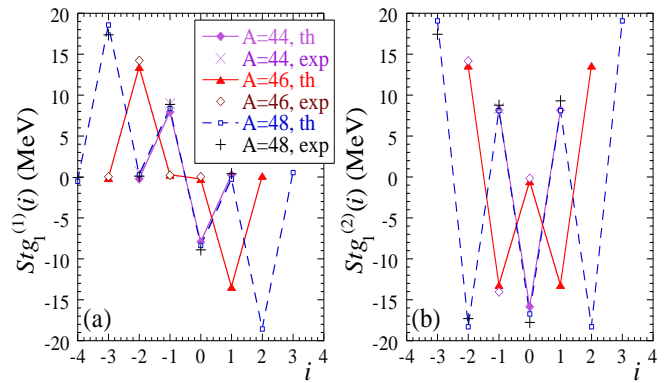,width=8.7cm,height=5.1cm}}}
\caption{(Color online) The $Stg^{(1,2)} _1 (i)$ discrete derivatives for different  isobaric
multiplets for even-$A$ nuclei with valence nucleons in the
$1f_{7/2}$ shell with
a core
$^{40}Ca$.}
\label{CaStg1_2i}
\end{figure}

\begin{figure}[th]
\centerline{\hbox{\epsfig{figure=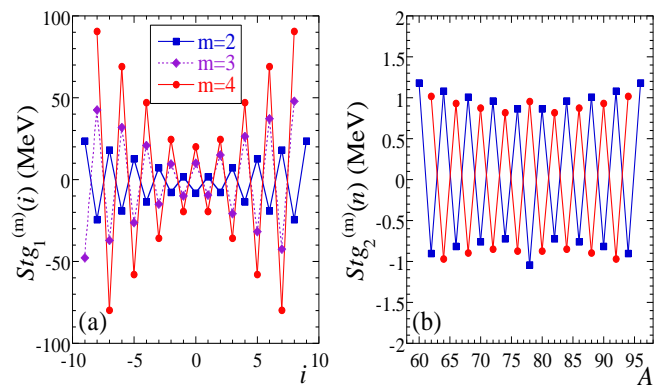,width=8.7cm,height=5.1cm}}}
\caption{(Color online) Discrete derivatives $Stg^{(m)} _\delta (i)$
($1f_{(\frac{5}{2})}2p_{(\frac{1}{2},\frac{3}{2})}1g_{(\frac{9}{2})}$,  a
$^{56}Ni$ core): (a)
$\delta =1$, $m=2,3,4$ for $A=76$ isobars; (b) $\delta =2$, $m=2,4$
for $(i=-1)$
multiplet [$N=Z+2$].}
\label{NiStg2_3_4i2_4n}
\end{figure}

\begin{figure}[th]
\centerline{\hbox{\epsfig{figure=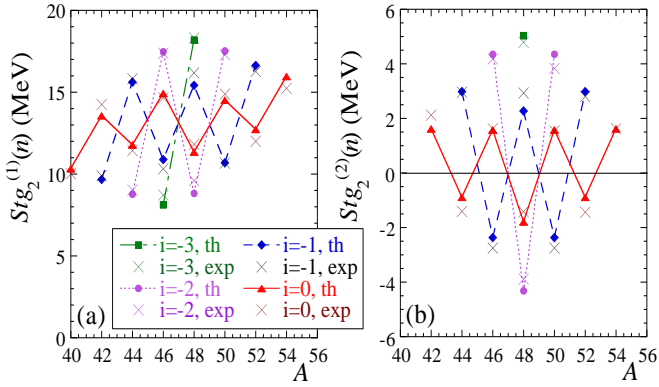,width=8.7cm,height=5.1cm}}}
\caption{(Color online) Discrete derivatives $Stg^{(1,2)} _2(n)$ for different
$i$ multiplets for even-$A$ nuclei ($1f_{7/2}$, a $^{40}Ca$ core).}
\label{CaStg1_2n}
\end{figure}

The discrete
derivatives, $Stg^{(m)} _1 (i)$, $m=1,2,...$, show a prominent
$\Delta i=1$  staggering of the experimental energies \cite{AudiWapstra} of the lowest
$0^{+}$ {\it
isovector-paired} states for different isobaric multiplets [see  Figure
\ref{CaStg1_2i} for the $1f_{7/2}$ shell and Figure
\ref{NiStg2_3_4i2_4n}(a) for
nuclei  above the $^{56}Ni$ core]. The theory reproduces this staggering very
well.
 For each of the $i$ multiplets ($i$ fixed), a
$\Delta n=2$ staggering effect is also observed for the experimental
values \cite{AudiWapstra} via
the energy filters $Stg^{(m)} _2 (n)$, $m=1,2,...$, and successfully predicted
by the symplectic model (Figure \ref{CaStg1_2n} ($1f_{7/2}$) and Figure
\ref{NiStg2_3_4i2_4n}(b)
($1f_{(\frac{5}{2})}2p_{(\frac{1}{2},\frac{3}{2})}1g_{(\frac{9}{2})}$)).

The staggering amplitudes of both $Stg^{(m)} _1 (i)$ and $Stg^{(m)}  _2 (n)$,
while almost independent of the total number of particles $n$, increase with
increasing difference in proton and neutron numbers, $i$, and  hence the
$ee-oo$ staggering effect is greater for the proton- (neutron-) rich nuclei
than around $N\approx Z$. Also, the amplitude of $Stg^{(m)} _1 (i)$
increases in
higher-order  derivatives. This  analysis shows a more complicated
dependence of
the energy function on the  isospin projection $i$  than on the mass
number $A$.

\begin{figure}[th]
\centerline {\epsfxsize=3.55in\epsfbox{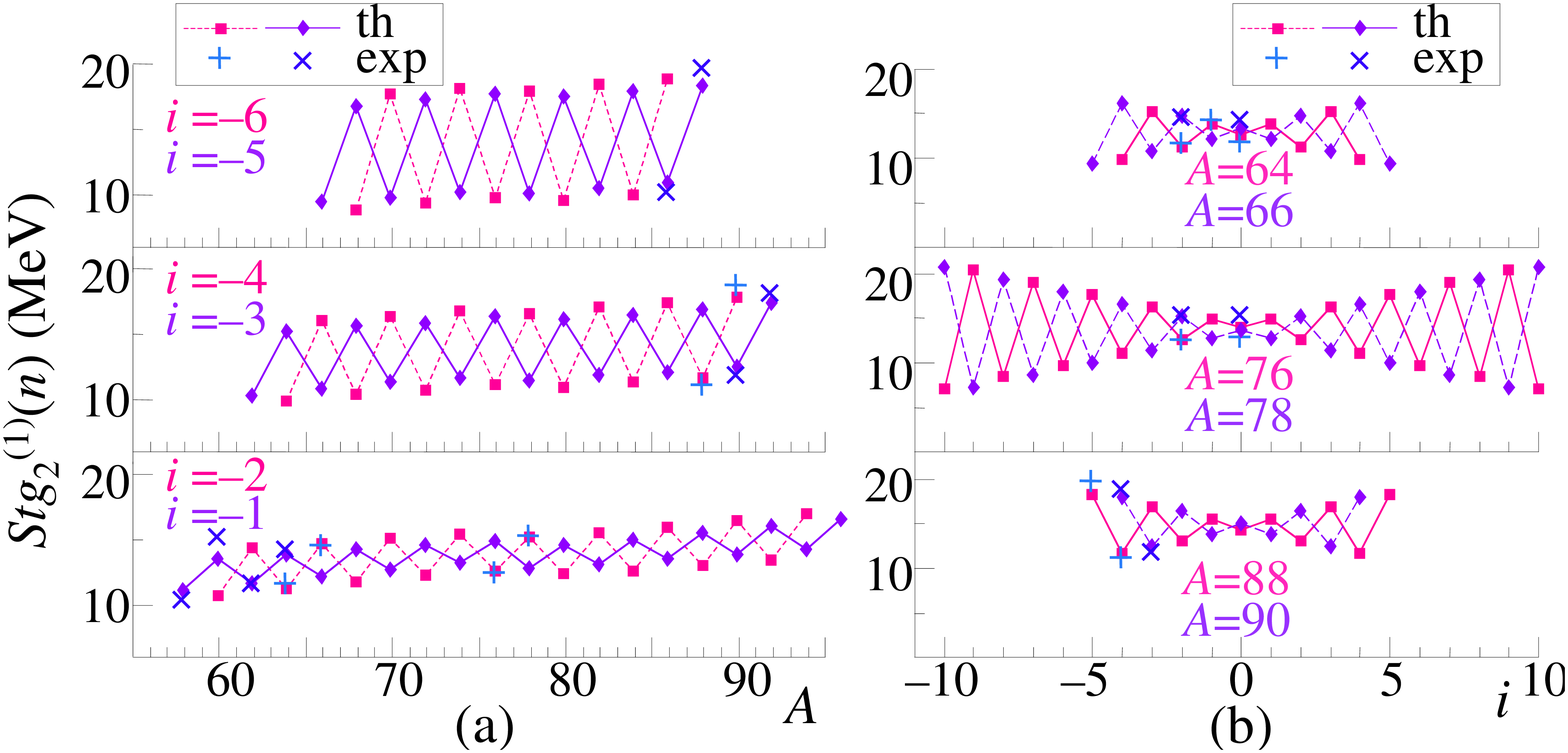} }
\caption{(Color online) The $Stg^{(1)} _2 (n)$ discrete approximation of the first
derivative,
$\partial E_0/\partial n$ ($^{56}Ni$ core) with respect to: (a) $A$ for several
$i$ multiplets; (b) $i$ for different isobars. }
\label{NiStg1n}
\end{figure}

The first, $m=1$, discrete derivative, $S_{pn}=2Stg^{(1)} _2
(n)=E_0(n+2)-E_0(n)$, where $i$ is fixed, corresponds to the energy
gained when a
$T=1$ $pn$ pair is  added [Figure
\ref{CaStg1_2n}(a) ($1f_{7/2}$) and Figure \ref{NiStg1n} (a $^{56}$Ni  core)].
$S_{pn}$ is the true
$pn$ separation energy only when $E_0$ is the binding energy of the  odd-odd
nucleus involved in its calculation. The experimental data, where available
\cite{AudiWapstra}, is also  shown in Figure \ref{NiStg1n} and the $Sp(4)$ model follows the
distinctive zigzag pattern very well. A
$\Delta n=4$ bifurcation separates the nuclei into two groups: one of
even-even
nuclei [($n/2+i$)-even] and another of odd-odd nuclei [($n/2+i$)-odd]. The
$S_{pn}$ energy difference has a smooth behavior within each group.
The magnitude
of
$S_{pn}$ is  proportional to the total number of particles and increases
(decreases) with $i$ for odd-odd  (even-even) nuclei (Figure
\ref{NiStg1n}) \footnote {When ($n/2+i$) corresponds to an odd-odd  nucleus
$S_{pn}$ is related to the properties of the even-even ($n+2$) nucleus.}.
Furthermore, the
$Stg^{(1)} _4 (n)=\frac{Stg^{(1)} _2 (n+2)+Stg^{(1)} _2 (n)}{2}$  energy
difference shows no
$\Delta n=4$ staggering (average values of two consecutive data
points in Figure
\ref{NiStg1n}). This indicates that the addition of an $\alpha $-like
cluster has
almost the same effect for both even-even and odd-odd nuclei. This
statement does
not  contradict the stronger binding of even-pairs nuclei as compared to
odd-pairs ones, which is  detected via $S_{pn}$ and the binding energy ($BE$) filter,
$BE(Z+2,N+2)-\frac{BE(Z+2,N)+BE(Z,N+2)}{2}$
\cite{GambhirRS83}.

\subsubsection{Pairing gaps}
The $Stg^{(m)} _1 (i)$ and $Stg^{(m)} _2 (n)$ energy differences,
$m=1,2,...$, described above, isolate effects related to the various  types of
pairing in addition to nonmonopole interactions resulting in changes in energy
due to the  different isospin values (symmetry term).
As noted in \cite{ZhangCB89,ZamfirC91}, the significance of the various energy
filters can be understood using phenomenological arguments  that can be given a
simple and useful graphical representation. Specifically, each nucleus can be
represented by an inactive core, schematically illustrated by a box, 
$\square $,
in which the interaction between the constituent  particles does not change.
Active particles beyond this core can be represented by solid or empty
dots, for protons or neutrons, above the box.

The second-order filter
\begin{eqnarray}
Stg^{(2)} _1 (i)&=&E_0(i+1)-2E_0(i)+E_0(i-1),\ n=\text{const}
\nonumber \\
&=&E_0(N_{+1}+1,N_{-1}-1)-2E_0(N_{+1},N_{-1}) \nonumber \\
&+&E_0(N_{+1}-1,N_{-1}+1)
\label{Stg2i}
\end{eqnarray}
when centered at an odd-odd [$(n/2+i)$-odd]
self-conjugate ($i=0$)
nucleus, represents the pairing gap relation $2\tilde{\Delta }$
\begin{eqnarray}
\hspace{-0.2in}
Stg^{(2)} _1 (i=0) &\stackrel{(\frac{n}{2}-odd)}{=}&
\stackrel{\bullet \bullet }{\square }+\stackrel{\circ \circ }{\square }-2
\stackrel{\bullet \circ  }{\square } \nonumber \\
&\approx &2\tilde{\Delta }\equiv 2\Delta _{pp}+2\Delta _{nn}-4\Delta _{pn}.
\label{Stg2iGap}
\end{eqnarray}
The result (\ref{Stg2iGap}) follows from the well-known definition  of the
empirical like-particle pairing gap \cite{BohrMottelson}
\begin{eqnarray}
&& \Delta _{pp(nn)}\equiv \nonumber \\
&& \equiv
\frac{1}{2}(BE(N_{+1}\pm 1,N_{-1}\mp 1)-BE(N_{+1}-1,N_{-1}-1) \nonumber \\
&& -2[BE(N_{\pm 1},N_{\mp 1}-1)-BE(N_{+1}-1,N_{-1}-1)]) \label{gapPPNN} \\
&&=\frac{1}{2}(\stackrel{\bullet \bullet }{\square }-\square -2[
\stackrel{\bullet }{\square }-\square]) \nonumber,
\end{eqnarray}
which isolates the isovector pairing interaction of the $(N_{\pm 1})^{th}$ and
$(N_{\pm 1}+1)^{th}$ protons (neutrons) for an even-even
($N_{+1}-1,N_{-1}-1$)-core  (marked by a square)
\cite{ZamfirC91}. We also define the $pn$ isovector pairing gap
\begin{eqnarray}
&&\Delta _{pn} \equiv \nonumber \\
&&\equiv \frac{1}{2}(E_0(N_{+1},N_{-1})-BE(N_{+1},N_{-1}-1) \nonumber \\
&&\ -[BE(N_{+1}-1,N_{-1})-BE(N_{+1}-1,N_{-1}-1)]) \nonumber \\
&&=\frac{1}{2}(\stackrel{\bullet \circ }{\square }-\stackrel{\bullet
}{\square }
-[ \stackrel{\circ }{\square }-\square])
\label{gapPN}
\end{eqnarray}
as the pairing interaction of the $(N_{+1})^{th}$ proton and the
$(N_{-1})^{th}$ neutron. In order to account correctly for the $T=1$
mode of the $pn$
pairing one should  consider in (\ref{gapPN}) the
$E_0$ energy of the odd-odd ($N_{+1},N_{-1}$) nucleus (that is, the
energy  of the
isobaric analog state rather than its ground state energy, $BE$). For the
remaining  even-even nuclei in (\ref{Stg2i}), replacing the symbol $E_0$ with
$BE$ is justified. In the computation  of $\tilde{\Delta }$, all
odd-$A$ binding
energies in (\ref{gapPPNN}) and (\ref{gapPN}) cancel so their  theoretical
calculation is not required.

\begin{figure}[h]
\centerline{
\hbox{\epsfig{figure=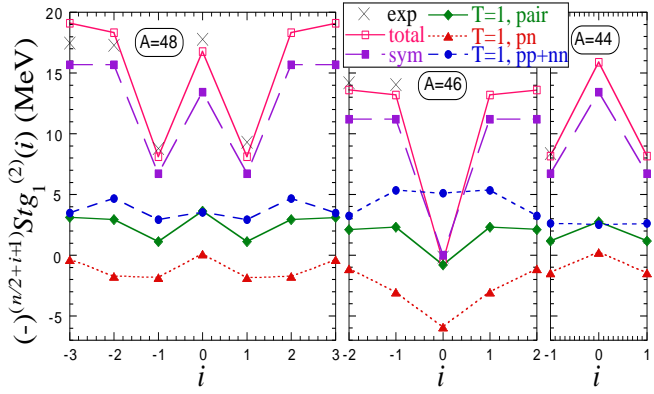,width=8.7cm,height=5.2cm}}}
\caption{(Color online) Theoretical staggering amplitudes for the total energy in
comparison to experiment \cite{AudiWapstra}, for the isovector pairing energy, the $pn$ and
the like-particle pairing  energies, and for the symmetry energy  for $A=48$, $A=46$ and
$A=44$ nuclei in the
$1f_{7/2}$ shell (a $^{40}Ca$ core).}
\label{CaHterms}
\end{figure}

The $\tilde{\Delta }$ relation of the gaps is a measure of the
difference in the
isovector pairing energy between even-even and odd-odd nuclei.  For
odd-odd $N=Z$
nuclei information about $\tilde{\Delta }$ is extracted via the $Stg^{(2)}  _1
(i)$ energy filter (\ref{Stg2i}). Both experimental and model estimations yield
$\tilde{\Delta }\cong 0$ for all the odd-odd $i=0$ nuclei in the
$1f_{7/2}$ shell
(for example, see  solid (purple) line with open squares in Figure
\ref{CaHterms}
for
$A=46$, $i=0$). The result reflects the  fact that in this case all three
isovector pairing gaps,
$\Delta_{pp}$, $\Delta_{nn}$ and
$\Delta_{pn}$, are equal \cite{Vogel00,Macchiavelli00}.

A different scenario regarding two aspects is encountered when one
considers the
$Stg^{(2)} _1 (i)$ discrete derivative centered at an even-($n/2+i$) $N=Z$
nucleus  [relative to a $(N_{+1}-2,N_{-1}-2)$-core]:
\begin{eqnarray} 
\hspace{-0.4cm}
Stg^{(2)} _1 (i=0)&=&
\stackrel{\stackrel{\bullet}{\bullet} \stackrel{\bullet }{\circ } }{\square
}+\stackrel{\stackrel{\circ }{\circ }
\stackrel{\bullet }{\circ }}{\square }-2
\stackrel{\stackrel{\bullet}{\bullet} \stackrel{\circ }{\circ }}{\square }\nonumber \\
&\approx &- \frac{2}{3}\tilde{\Delta }+I_2^{J\neq 0,T\neq 1},\ \
(\frac{n}{2}+i)\text{ even},
\label{Stg2i_int_i0}
\end{eqnarray} where an additional nonpairing two-body interaction
$I_2^{J\neq
0,T\neq 1}$ is not filtered out in this case. Here, for example, $I_2^{J\neq
0,T\neq 1}$ is related  to the nonpairing interaction of the three protons and
of the three neutrons in the odd-odd nuclei  (\ref{Stg2i_int_i0}). Another new
feature of (\ref{Stg2i_int_i0}) is that
$Stg^{(2)} _1 (i=0)$ does not simply account for the energy gained when two
$pn$ pairs are created (in the first two odd-odd nuclei) and the
energy lost to
destroy a $pp$ pair and a $nn$ pair in the even-even $N=Z$ nucleus. The
straightforward  reason is that $pp$, $nn$ and
$pn$ $T=1$ pairs coexist. A good approximation that serves well in  estimating
the pairing gaps is to assume that a $2p-2n$ formation above the inactive core
($\square $) consists of $n_0=2/3$ $pn$ pairs, $n_{ 1}=2/3$ $pp$ pairs and
$n_{- 1}=2/3$ $nn$ pairs [rather than a proton pair ($n_1=1$) and a
neutron pair
($n_{-1}=1$)]. This is in analogy to an even-even $n=4$ nucleus where the $pp$,
$nn$ and
$pn$ ``numbers of pairs" are the same and equal to one third the
total number of
pairs, $n/2$
\cite{EngelLangankeVogelDobes,SGD03}. Additionally, the relations like
(\ref{Stg2iGap}) -  (\ref{Stg2i_int_i0}) are based on the assumptions that the
interaction of a particle with the core is independent of  the type of
added/removed particles and is the same for all protons (neutrons) above the
core.  Finally, all the approximations are of an order $O(1/\Omega)$.

The additional nonmonopole two-body residual interaction $I_2^{J\neq
0,T\neq 1}$
should be also taken into account for the rest $i\neq 0$ of the ($ee$,  and $oo$) nuclei
\begin{eqnarray} Stg^{(2)} _1 (i\neq 0)\approx \left\{
\begin{array}{ll}
    -  \frac{4}{3}\tilde{\Delta }+I_2^{J\neq 0,T\neq 1} &, ee \\
       \frac{4}{3}\tilde{\Delta }+I_2^{J\neq 0,T\neq 1} &, oo.
\end{array}
\right.
\label{Stg2i_int}
\end{eqnarray} The main contribution to the $I_2^{J\neq 0,T\neq 1}$ interaction
is  due to the symmetry energy as is apparent from the $Sp(4)$ model.

The very close theoretical reproduction of the experimental
staggering allows us
to use the symplectic model as a microscopic explanation of the
observed effects
through the investigation of the different terms in the Hamiltonian
(\ref{clH})
(Figure \ref{CaHterms}). According to the $Sp(4)$ model, the $ee-oo$
staggering patterns  appear due to the discontinuous change of the seniority
numbers driven by the $T=1$ pairing  interaction \cite{SGD03}.
Even values of the seniority quantum number ($\nu _1$) in even-even nuclei and
odd values for odd-odd nuclei lead to a change in $pn$ and 
like-particle pairing
energies in opposite directions. After the contribution from the isovector
pairing energy is taken away, the  theoretical staggering amplitude,
$(-)^{\frac{n}{2}+i+1}Stg^{(2)} _1 (i)$, has still a (typically
large) component from the remaining
$(J\neq 0,T\neq 1)$ interactions in the Hamiltonian (\ref{clH}),  mainly the
symmetry ($T^2$) term [Figure \ref{CaHterms}, long-dashed (purple) line with
squares]. This  is the same nonmonopole nuclear interaction,
$I_2^{J\neq 0,T\neq
1}$, that was suggested in  (\ref{Stg2i_int_i0}) and (\ref{Stg2i_int}) using
phenomenological arguments. Indeed, the  symmetry energy contribution is
significant and nonzero in all nuclei but the odd-odd $N=Z$
(\ref{clH}) (Figure
\ref{CaHterms}), which is consistent with the discussion above
[(\ref{Stg2iGap}),  (\ref{Stg2i_int_i0}), (\ref{Stg2i_int})]. Also,
an estimation
of the pairing gaps is  possible based on the examination of the model
Hamiltonian but the theoretical staggering amplitudes of  the $T=1$ pairing
energies (shown in Figure \ref{CaHterms}) need to be rescaled in
accordance with
(\ref{Stg2iGap}), (\ref{Stg2i_int_i0}) and (\ref{Stg2i_int}).

\begin{figure}[h]
\centerline {\epsfxsize=3.55in\epsfbox{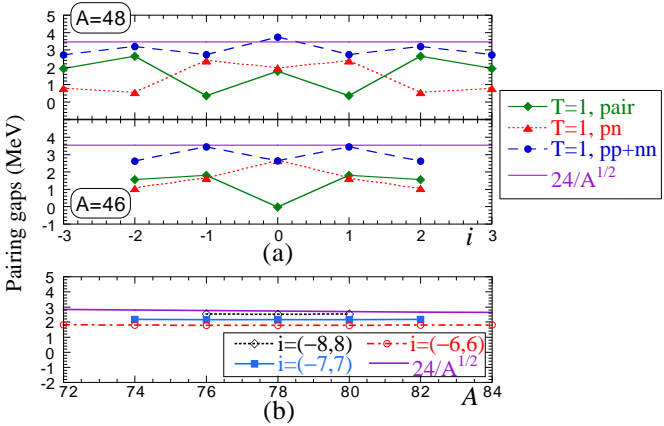} }
\caption{(Color online) Estimation of the pairing gaps: (a) total isovector pairing  gap
$\tilde{\Delta }$, $2\Delta _{pn}$ and $\Delta _{pp}+\Delta _{nn}$,
as well as the
empirical  like-particle pairing gap $\Delta _{pp}+\Delta
_{nn}=24/A^{1/2}$ shown
for comparison, for $A=48$ and
$A=46$ nuclei versus $i$ ($1f_{7/2}$ shell); (b) like-particle pairing gap
(according to  (\ref{Stg4i_int})) versus $A$ for
$i=\pm 6, \pm 7, \pm 8$ multiplets in the
$1f_{(\frac{5}{2})}2p_{(\frac{1}{2},\frac{3}{2})}1g_{(\frac{9}{2})}$ shell.}
\label{CaNigaps}
\end{figure}

In a way analogous to that used in (\ref{Stg2i_int}), the  second-order
discrete derivative with respect to
$n$ (can be compared to the filter used in \cite{Macchiavelli00})
\begin{eqnarray} Stg^{(2)} _2(n)&=\frac{E_0(n+2)-2E_0(n)+E_0(n-2)}{4},\ i=\text{const}
\label{Stg2n}
\end{eqnarray} is related to the pairing gap relation
\begin{eqnarray} Stg^{(2)} _2 (n)\approx  \left\{
\begin{array}{ll}
    -\frac{\tilde{\Delta }}{3}+I_2^{J\neq 0,T\neq 1} &, ee \\
     \frac{\tilde{\Delta }}{3}+I_2^{J\neq 0,T\neq 1} &, oo,
\end{array}
\right.
\label{Stg2n_int}
\end{eqnarray} where in the odd-odd case, for example, $I_2^{J\neq 0,T\neq 1}$
is  the non-pair interaction of the last two protons with the last two neutrons
in the ($n+2$) nucleus.  The effects due to $\tilde{\Delta }$ cannot
be isolated
via (\ref{Stg2n_int}) because of the additional  nonzero
contribution due to the
symmetry energy. However, the staggering amplitude of the  discrete derivative
(\ref{Stg2n}),
$-3(-)^{\frac{n}{2}+i}Stg^{(2)} _2 (n)$, of the  theoretical total,
$pp$ ($nn$) and $pn$ pairing energies can provide for estimation of
the pairing
gaps
$\tilde{\Delta }$, $\Delta _{pp(nn)}$ and $-2\Delta _{pn}$,  respectively
[Figure \ref{CaNigaps}(a)]. The like-particle pairing gap can be
compared to the
empirical value of
$\Delta _{pp}+\Delta _{nn}=24/A^{1/2}$ \cite{BohrMottelson} [solid  (purple)
line]. The gap is smaller in odd-odd nuclei as compared to their even-even
neighbors. This is a  consequence of a decrease in the like-particle pairing
energy in the odd-odd nuclei due to the  blocking effect while there is an
increase in energy due to the $pn$ pairing. The $pn$ isovector  pairing gap
increases toward $i=0$ and eventually gets almost equal to $\Delta
_{pp(nn)}$ for
odd-odd nuclei  around the $N=Z$ region, which is in agreement with the
discussion of \cite{Macchiavelli00,Vogel00}.

Furthermore, an average of the additional non-pair interaction is
achieved by the
fourth-order derivatives both in $n$ [$Stg^{(4)} _2 (n)$] and $i$
[$Stg^{(4)} _1(i)$]
\begin{eqnarray}
\tilde{\Delta}_{|i|\neq 0,1} &\approx &\frac{3}{16}(-)^{n/2+i}(Stg^{(4)} _1
(i)-I_2^{J\neq 0,T\neq  1}) \label{Stg4i_int} \\ &\approx
&3(-)^{n/2+i}(Stg^{(4)}
_2 (n)-I_2^{J\neq 0,T\neq 1}).
\label{Stg4n_int}
\end{eqnarray} Assuming that the $pn$ pairing gap is negligible for high-$i$
nuclei  in large shells like the
$1f_{(\frac{5}{2})}2p_{(\frac{1}{2},\frac{3}{2})}1g_{(\frac{9}{2})}$ 
major shell, the gap relation
(\ref{Stg4i_int}) or (\ref{Stg4n_int}) provides for a rough estimation of the
like-particle pairing gaps. With the use of the model Hamiltonian
(\ref{clH}) we
can estimate the additional
$I_2^{J\neq 0,T\neq 1}$ interaction with the major input being the symmetry
energy. Although the  existence of a very small mixing of isospin values
complicates the computation of the symmetry energy for  nuclear
systems with very
large interaction matrices, as a very good approximation one may use
$E_{sym,T}=\frac{E}{2\Omega }T(T+1)$ with isospin values $T =|i|$ for even-even
nuclei and $T =|i|+1$ for  odd-odd nuclei. Once the fourth-order discrete
derivative (\ref{Stag_m}) of the  approximated symmetry energy is removed from
$Stg^{(4)} _1 (i)$ (\ref{Stg4i_int}), the like-particle pairing gaps $\Delta
_{pp}+\Delta _{nn}$ are found to be in a very  good agreement with the
experimental approximation of
$24/\sqrt{A}$ for the ($i=\pm 6,\ \pm 7,\ \pm  8$) multiplets in the
$1f_{(\frac{5}{2})}2p_{(\frac{1}{2},\frac{3}{2})}1g_{(\frac{9}{2})}$ major
shell (Figure \ref{CaNigaps}(b)). For lower $|i|$ values the 
difference increases
due to an increase in the $pn$ pairing gap  as mentioned above. As a whole, the
agreement would not be possible if the significant energy  contribution due to
the symmetry energy was not taken into account.

\subsubsection{Second-order mixed derivatives}

Next we consider the second-order discrete mixed derivative of the  relevant
energies with respect to the total number $n$ and the third projection $i$
\begin{eqnarray} &Stg^{(2)} _{2,1} (n,i)=
\nonumber \\
=& \frac{E_0(n+2,i+1)-E_0(n+2,i)-E_0(n,i+1)+E_0(n,i)}{2}
\label{Stg2ni} \\
\approx &\left\{
\begin{array}{ll}
      \frac{2}{3}\tilde{\Delta }+I_2^{J\neq 0,T\neq 1} &, ee \\
     -\frac{2}{3}\tilde{\Delta }+I_2^{J\neq 0,T\neq 1} &, oo,
\end{array}
\right.
\label{Stg2ni_int}
\end{eqnarray} 
where in addition to the pairing gaps relation,
$\tilde{\Delta }$,
there is the contribution due to the nonpairing interaction,
$I_2^{J\neq 0,T\neq
1}$. For  example, for the odd-odd \{even-even\} case it is the
positive \{negative\}
nonpairing average  interaction between the last three protons
\{neutrons\} in the
($n+2$ $\{n\},i+1$) nucleus with a  ($n-2\ \{n-4\},i$) core. Within the
$Sp(4)$ framework
the additional nonpairing contribution corresponds to the  staggering of the
symmetry energy approximation, $E_{sym,T}$, of $(-)^{n/2+i+1} \frac{E}{2\Omega
}(2|i|+3)$.
\begin{figure}[th]
\centerline{\epsfxsize=3.55in\epsfbox{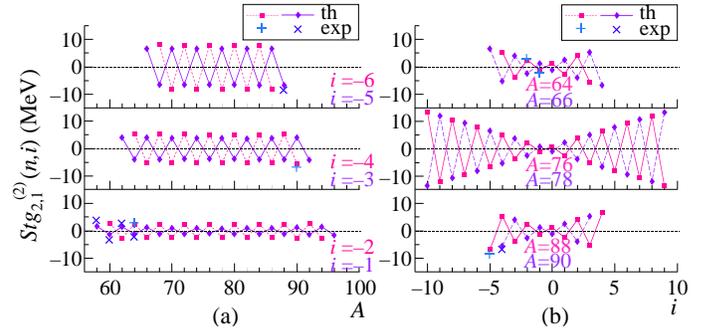}}
\caption{(Color online) Second-order energy filter
$Stg^{(2)} _{2,1} (n,i)$  for nuclei above the $^{56}Ni$ core with  respect to
$A$ (a) and $i$ (b).}
\label{derNT}
\end{figure}

The filter (\ref{Stg2ni}) isolates fine structure effects between two
$i$ multiplets [Figure~\ref{derNT}(a)] and two consecutive isobaric sequences
[Figure~\ref{derNT}(b)]. Clearly, it reveals a
$\{\Delta n,\Delta i\}=\{2,1\}$ symmetric oscillating pattern as it
is observed in the experiment \cite{AudiWapstra}. Its positive (negative) value is centered
at even-even (odd-odd)  nuclei and its amplitude increases (decreases) with $|i|$.
This mixed
discrete derivative  (\ref{Stg2ni}) serves as another test for the
$Sp(4)$ model
and allows for a detailed  investigation of the nonpairing, like-particle
interactions involved.

\begin{figure}[h]
\centerline{
\hbox{\epsfig{figure=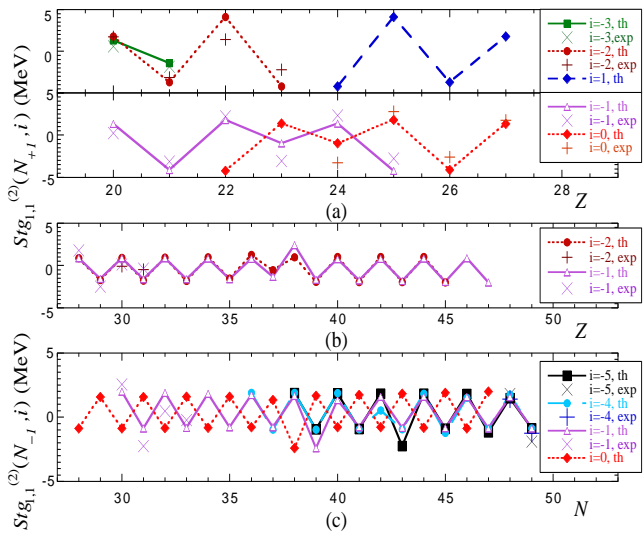,width=8.7cm,height=7cm}}}
\caption{(Color online) Discrete derivative, $Stg^{(2)} _{1,1} (x,i)$, for various
$i$ multiplets for even-$A$ nuclei: (a) $x=N_{+1}$, $1f_{7/2}$
level; (b) $x=N_{+1}$,
$1f_{(\frac{5}{2})}2p_{(\frac{1}{2},\frac{3}{2})}1g_{(\frac{9}{2})}$ shell; (c)
$x=N_{-1}$,
$1f_{(\frac{5}{2})}2p_{(\frac{1}{2},\frac{3}{2})}1g_{(\frac{9}{2})}$ shell.}
\label{NiderNpTau0}
\end{figure}

To isolate the effect of nonpairing interactions (again, it is  understood to
order
$1/\Omega$), an energy difference with respect to both
$N_{\pm 1}$ and $i$ can be considered. The second discrete derivative  of the
energy
\begin{eqnarray} &Stg^{(2)} _{1,1} (N_{\pm 1},i)= \nonumber \\
=&\frac{E_0(N_{\pm 1}+1,i+1)-E_0(N_{\pm 1}+1,i)-E_0(N_{\pm
1},i+1)+E_0(N_{\pm 1},i)}{2}
\label{Stg2npi}
\end{eqnarray} represents the negative \{positive\} nonpairing two-body
interaction  of the last two neutrons \{protons\} with a proton and a neutron in
the ($N_{\pm 1}+1,i\ \{+1\}$)  nucleus. It shows prominent
$\Delta i=1$ staggering patterns for different $i$ multiplets
(Figure~\ref{NiderNpTau0}). While in the framework of the $Sp(4)$ model its
amplitude does not depend on
$N_{\pm 1}$ and $i$ except for irregularities around the mid-shell,
the magnitude of the few experimental values \cite{AudiWapstra} (where data exist) tends to
be slightly lower away from the  closed shell. As a whole, the results show that the
staggering behavior of this interaction is due  to the fine structure features in the
relationship between the like-particle and $pn$ nonpairing  interactions and
differs between proton-rich and neutron-rich nuclei.

Regarding (\ref{Stg2npi}) and the other discrete approximations of  the
derivatives in Section
\ref{DerStg}, it is clear that the oscillating patterns that exist  and their
regular appearance throughout the nuclear chart cannot be a simple artifact due
to  errors in the experimental or theoretical energies. Even more,
the staggering
amplitudes are  usually (very) large  compared to the energy uncertainties.

For all the discrete derivatives that we have investigated above and that show
``$ee$-$oo$" staggering behavior, the discontinuity of the symmetry
term (due to
discrete changes in the isospin value) plays an important role. In
contrast, when
these  discrete derivatives include states of odd-odd nuclei with a
dominant $T=0$
$pn$ coupling there is  a constant or no contribution due to the
symmetry energy,
and hence yield patterns of different  shapes and interpretations. Our
investigation does not aim to account for such effects. It is focused  on the
$ee-oo$ staggering behavior of the $E_0$ energies of the lowest {\it
isovector-paired}  states as observed from the experimental data and reproduced
remarkably well by the $Sp(4)$ model.

\section{Conclusions}
A dynamical $Sp(4)$ symmetry was used to provide for a
natural  classification scheme of nuclei and to describe isovector pairing
correlations and high-$J$ interactions. In  a previous study \cite{SGD03},
it was found that the $Sp(4)$ model reproduced reasonably well the
experimental energies of the lowest {\it isovector-paired}
$0^+$ states and provided for an  estimation of the interaction strength
parameters.

Here the $sp(4)$ algebraic approach has been further tested through
second- and higher-order discrete derivatives of the energies of the 
lowest {\it
isovector-paired}  $0^+$ states in the $Sp(4)$ systematics, without
any parameter variation. If reality were only a  mean-field theory, none of the
finite energy differences would reveal regular or irregular 
staggering  effects.
The reason is that any effect due to a smoothly varying mean-field part of
the nuclear interaction is either entirely cancelled out in a finite energy
difference filter or contributes regularly to the isolated part of the
interaction. Indeed, the results obtained show that this is not the case
and staggering behavior is observed. The theoretical discrete derivatives
investigated not only followed the experimental patterns but their 
magnitude was
found to be in a  remarkable agreement with the data.  The proposed model
successfully interpreted the following: the two-proton (two-neutron)  separation energy
$S_{2p(2n)}$ (hence determined the two-proton drip line) for even-even nuclei,
the  $S_{pn}$ energy difference when a
$pn$ $T=1$ pair is added, the observed irregularities around $N=Z$,
the prominent
$ee-oo$ staggering when even-even and odd-odd nuclides are considered
simultaneously, the like-particle and
$pn$ isovector pairing gaps, and the large contribution to the finite  energy
differences due to the symmetry term. The oscillating effects, where observed,
were found to  develop due to the discontinuity of the seniority
numbers for the
$pn$ and like-particle  isovector pairing, which is in addition to the larger
staggering due to the  discontinuous change in isospin values (symmetry term)
between even-even and odd-odd nuclei.

We found a finite energy difference that, for a specific case, can be
interpreted as an isovector pairing gap, $\tilde{\Delta
}=\Delta_{pp}+\Delta_{nn}-2\Delta_{pn}$,  which is related to the like-particle
and $pn$ isovector pairing gaps. They correspond to the
$T=1$ pairing mode because we do not consider the binding energies for all the
nuclei  but the respective isobaric analog
$0^+$ states for the odd-odd nuclei with a $J\neq 0^+$ ground state.  This
investigation is the first of its kind. Moreover, the relevant energies are
corrected  for the Coulomb interaction and therefore the isolated effects
reflect solely the nature of the  nuclear interaction.

The outcome of this investigation shows that, in comparison to the experiment, the
simple $Sp(4)$ model reproduces not only global trends of the relevant energies
but as  well the smaller fine structure effects driven by isovector pairing
correlations and higher-$J$ $pn$  and like-particle nuclear interactions.
In particular, the $sp(4)$ algebraic model was used to interpret  specific
phenomena revealed in finite energy differences and to investigate the
contribution of the underlying interactions. In this  way, it provides
for an estimation of the isovector pairing gaps. For $N=Z$ odd-odd nuclei all
three  pairing gaps were found equal while the $pn$ pairing was found to weaken
relative to the like-particle pairing  strengths with increasing proton
(neutron) excess. The like-particle pairing gaps were found to be in  a good
agreement with the empirical value of
$12/\sqrt{A}$. Additionally, the discrete derivatives give  insight into
particular small parts of the various non-($J=0,T=1$) interactions, mainly into
the detailed  contribution of the interaction related to the $T(T+1)$ term
(symmetry energy). Small deviations from the experimental data are
attributed to
other two-body interactions or higher-order correlations that are not  included
in the theoretical model.

We explored independent finite energy differences based on a simple
$sp(4)$ algebraic classification scheme. The results suggest that
this theoretical
framework can be  used to reproduce various experimental results including
observed staggering behavior in fine structure  effects of nuclear collective
motion.

\vskip 0.5cm
\noindent {\bf Acknowledgments}

This work was supported by the US National Science Foundation through
Grant No.
0140300.  K.D.S. acknowledges supplemental support from the Graduate School of
Louisiana State University.

\end{document}